\documentclass[fleqn,usenatbib]{mnras}

\usepackage{newtxtext,newtxmath}

\usepackage[T1]{fontenc}
\usepackage{ae,aecompl}
\usepackage{adjustbox}
\usepackage{placeins}
\usepackage{enumitem}
\usepackage{caption}
\usepackage{comment}
\usepackage{float}
\usepackage{subcaption}
\usepackage{hyperref}
\usepackage{graphicx}
\usepackage{multirow}
\usepackage{booktabs}
\newcommand{\RNum}[1]{\uppercase\expandafter{\romannumeral #1\relax}}

\usepackage{graphicx}	
\usepackage{amsmath}	
\usepackage{amssymb}	

\title{Timing of young radio pulsars $\rm{\RNum{2}}$. Braking indices and their interpretation}

\author[A. Parthasarathy et al.]
{A. Parthasarathy$^{1,2,3,4}$,\thanks{E-mail: aparthas@mpifr-bonn.mpg.de}
S. Johnston$^{3}$,
R.M. Shannon$^{1,2}$,
L. Lentati$^{5}$,
M. Bailes$^{1,2}$, \newauthor 
S. Dai$^{3}$, 
M. Kerr$^{6}$, 
R.N. Manchester$^{3}$,
S. Os{\l}owski$^{1,2}$,
C. Sobey$^{7}$, \newauthor
W. van Straten$^{8}$,
P. Weltevrede$^{9}$
\\
$^{1}$ Centre for Astrophysics and Supercomputing, Swinburne University of Technology, P.O. Box 218, Hawthorn, Victoria 3122, Australia \\
$^{2}$ OzGrav: Australian Research Council Centre of Excellence for Gravitational Wave Discovery.  \\
$^{3}$ CSIRO Astronomy and Space Science, Australia Telescope National Facility, PO~Box~76, Epping NSW~1710, Australia \\
$^{4}$ Max-Planck-Institut f\"{u}r Radioastronomie, Auf dem H\"{u}gel 69, D-53121 Bonn, Germany \\
$^{5}$ Astrophysics Group, Cavendish Laboratory, JJ Thomson Avenue,  Cambridge, CB3 0HE, UK \\
$^{6}$ Space Science Division, Naval Research Laboratory, Washington, DC 20375, USA \\
$^{7}$ CSIRO Astronomy and Space Science, PO Box 1130 Bentley, WA 6102, Australia \\
$^{8}$ Institute for Radio Astronomy \& Space Research, Auckland University of Technology, Private Bag 92006, Auckland 1142, New Zealand \\
$^{9}$ Jodrell Bank Centre for Astrophysics, The University of Manchester, Alan Turing Building, Manchester, M13 9PL, United Kingdom \\ 
}
\date{Accepted XXX. Received YYY; in original form ZZZ}

\pubyear{2019}

\begin{document}
\label{firstpage}
\pagerange{\pageref{firstpage}--\pageref{lastpage}}
\maketitle

\begin{abstract}
In Paper I of this series, we detected a significant value of the braking index ($n$) for 19 young, high-$\dot{E}$ radio pulsars using $\sim$ 10 years of timing observations from the 64-m Parkes radio telescope. Here we investigate this result in more detail using a Bayesian pulsar timing framework to model timing noise and to perform selection to distinguish between models containing exponential glitch recovery and braking index signatures. We show that consistent values of $n$ are maintained with the addition of substantial archival data, even in the presence of glitches. We provide strong arguments that our measurements are unlikely due to exponential recovery signals from unseen glitches even though glitches play a key role in the evolution of a pulsar's spin frequency. We conclude that, at least over decadal time scales, the value of $n$ can be significantly larger than the canonical 3 and discuss the implications for the evolution of pulsars.


\end{abstract}


\graphicspath{ {./plots/} }



\section{Introduction}
The spin-down of pulsars is one of the most prominent features of their rotation (\citealt{davies_spindown}, \citealt{cole_spindown}). Considering the magnetic field of a neutron star in a vacuum to have dipolar configuration and assuming that the magnetic axis is misaligned with the rotation axis, it must radiate and lose energy, which contributes to the observed slow down (\citealt{gunn_ostriker}). Under the simplest assumption that magnetic dipole radiation is the only braking mechanism, we expect the spin-frequency ($\nu$) and the spin-frequency derivative ($\dot{\nu}$) to follow, $\dot{\nu}$ $\propto$ $-\nu^3 $. However, pulsars reside in dense plasma environments in which particles are accelerated up to very high energies. Such plasma outflows can result in radial deformation of pulsar's magnetic field lines, which could result in a large fraction of the angular momentum and energy loss in the form of particles (\citealt{precise_BI}, \citealt{harding_1999}), and additionally contribute to the underlying magnetic braking (\citealt{spitkovsky1}). The observations of quasi-periodic magnetospheric processes correlated with spin-down measurements (\citealt{kramer_magnetosphere}, \citealt{lyne_switched_magnetosphere}), strengthens the argument that a pulsar wind plays a substantial role in the pulsar braking mechanism. Observations of glitches, sudden spin-up events, alter the normal spin-down of pulsars (\citealt{espinoza_305}). Glitches are thought to arise from the transfer of angular momentum from a more rapidly spinning neutron star superfluid component to the crust (\citealt{baym_1969}, \citealt{anderson_itoh}, \citealt{melatos_glitch}). Coupling and decoupling between the crust and the superfluid core can decrease the moment of inertia with time, leading to further departures from the conventional spin-down due to magnetic dipole braking (\citealt{Andersson_Ho}). Pulsar braking is also thought to be dependent on the temporal evolution of the magnetic field strength ($B$) and the inclination angle between the magnetic and the rotational axis ($\alpha$) (\citealt{tauris}, \citealt{Young_2009}, \citealt{lyne_2013}, \citealt{JK17}). Finally, gravitational wave emission can also contribute to the energy losses in a pulsar. Thus, the supposition that a neutron star is a magnetic dipole rotating in a vacuum must be treated with caution. 

The braking index ($n$) is used to parameterize the spin evolution of pulsars and is defined by relating the spin-frequency and its derivative through a power law:
  \begin{equation}
  \label{pulsar_spindown}
  \dot{\nu} = -K\nu^{\rm n}.
  \end{equation}

$K$ depends on the moment of inertia of the neutron star and its magnetic moment. Although $K$ is often assumed to be constant in time, deviations from a constant magnetic moment can occur due to decay (or growth) of the magnetic field (\citealt{B_growth}) or changes in the dipole alignment (\citealt{dipole_growth}) which makes $K$ time-dependent (\citealt{Blandford_Romani}, \citealt{mckinney}).  
Neglecting these effects allows us to write the braking index as,
  \begin{equation}
  \label{braking_index}
  n = \frac{\nu\ddot{\nu}}{\dot{\nu}^2},
  \end{equation}
where $\ddot{\nu}$ is the second derivative of the spin-frequency. 

The characteristic age of the pulsar is usually written as $\tau_c = -\nu/2\dot{\nu}$ with the implicit assumption that $n=3$ and that the initial period of the pulsar is much less than its current period. However, given the variety of phenomena discussed above that can potentially contribute to pulsar braking, the characteristic age must be taken with caution (when interpreting it as a bona-fide age estimate of the pulsar). 
 
There are challenges associated with measuring the long-term $\ddot{\nu}$. In most cases, the value arising from pure spin-down is dominated by other phenomena like timing noise or glitches. Timing noise is caused by stochastic rotational irregularities of the neutron star about a steady state (\citealt{groth_timingnoise}) and is seen in all classes of pulsars (\citealt{hobbs_whitening}, \citealt{shannon_timingnoise},  \citealt{namkham_timingnoise}, \citealt{psj1}). The cause of timing noise has been attributed to both neutron star interiors, arising from the cross-coupling between the neutron star crust and its superfluid core (\citealt{superfluid_timingnoise}), or from magnetospheric torque fluctuations (\citealt{glitches_timingnoise}). Failure to model timing noise leads to biased estimates of pulsar parameters (\citealt{coles_whitening}). Developing upon previous Bayesian pulsar timing frameworks (\citealt{tn_haasteren}, \citealt{temponest}), \cite{psj1} (hereafter PSJ19) characterise the timing noise as a power-law process (see Equation \ref{red_noise}) in a sample of 85 high $\dot{E}$, young radio pulsars and report the median red-noise amplitude, $\log_{\rm {10}} \left(\frac{A_{\rm red}}{yr^{3/2}}\right)$ to be $-10.4^{+1.8}_{-1.7}$ and the spectral index ($\beta$) to be $-5.2^{+3.0}_{-3.8}$ and show that the strength of timing noise scales proportionally to $\nu^{1}|\dot{\nu}|^{-0.6\pm0.1}$. The modelling of timing noise simultaneously with the timing model will allow for unbiased measurements of pulsar parameters.

The presence of glitches, and in particular the glitch recovery, further complicates the measurement of $\ddot{\nu}$ and hence the braking index. Following a glitch event, a pulsar enters a recovery stage during which an increase in the spin-down rate and a relaxation towards the pre-glitch rotational state is observed over a timescale of days to years (\citealt{yu_glitches}, \citealt{fuentes_glitch}). Both the spin-up event and the subsequent recovery stage are interpreted as the presence of a superfluid component in the inner crust and core of the star (\citealt{haskell_melatos_glitches}). \cite{johnstongalloway_braking1} and \cite{alpar_baykal} point out that a post-glitch relaxation process can dominate the spin-down evolution of a pulsar resulting in large ($n$ > 3) values of braking indices. 

Because of the difficulties outlined above, it has been common in the literature to only report values of $n$ when they are close to the canonical value of 3. \cite{Lyne_CrabL} reported a braking index of 2.35(1) for the Crab pulsar (PSR J0534+2200) from 45 years of timing data. They note that the braking index value is close to 2.5 in the intervals between glitches. Similarly, the braking index and spin-down properties of the Vela pulsar are also well studied. While \cite{Shannon_Vela} determined the long-term value of $n$ to be $<8$, \cite{Espinoza_Vela} reported $n=1.7(2)$ using different techniques. Using inter-glitch models to obtain spin-down rates, \cite{Akbal_2017_vela} measured $n=2.8(1)$. It is important to note the reported long-term values of $n$ are smaller than the values reported for the linear regimes between glitches (where $n\sim30$). 

Braking index values ranging from $\sim0.03$ for PSR~B0540--69 (\citealt{Marshal_lowBI}) to $\sim3.15$ for PSR~J1640--4631 (\citealt{Archibald_highBI} have been reported. In PSR~J1513--5908, the braking index was measured in a series of papers (\citealt{Kaspi_1513}, \citealt{Livingstone_1513}, \citealt{Livingstone_kaspi_2011}) with the updated value being $2.832\pm0.003$ . This pulsar has never been observed to glitch in $\sim25$ years of observations. In contrast, PSR~J0537--6910 has the highest glitch rate of any pulsar (\citealt{middleditch_0547}) with $\sim$45 observed glitches in $\sim14$ years of observations. \cite{0537_negativeBI} reported that the spin-evolution of this pulsar is characterised by a well-defined negative braking index of $-1.22(4)$. Using the same dataset, \cite{ferdman_glitch_bi} and \cite{Andersson_2018_Braking} reported that the trend in the interglitch behaviour leads to a larger value of $n\sim7$.

In this paper, we measure $n$ for a sample of young pulsars using Bayesian inference. In Section \ref{measured_BI}, we report the measured values of $n$ for the 19 pulsars in our sample. In Section \ref{robustness_BI}, we use simulated and historical data sets to test the robustness of the measured values. In Section \ref{glitch_recovery_BI} we introduce a glitch recovery model to test if unseen glitches before our data set affect the measurement of $n$. In Section \ref{glitch_sim_sec}, using Monte Carlo simulations we investigate the presence of exponential glitch recoveries on long timescales. In Section \ref{implications_sec}, we develop a parameterization for the detectability of braking index and discuss the implications of high values of $n$ and the evolution of pulsars with time.

\section{Measured Braking Indices} \label{measured_BI}
In PSJ19, we described the observations and the methodology involved in obtaining timing solutions for 85 young radio pulsars observed with the CSIRO 64-m Parkes radio telescope. A total of 19 pulsars had a preferred model which included $\ddot{\nu}$ as shown in Table~3 of PSJ19. Here we reproduce the values of $\log_{\rm 10}(A_{\rm red})$, $\beta$, $\nu$, $\dot{\nu}$, $\ddot{\nu}$, $n$ and the preferred model for these 19 pulsars in Table \ref{braking_index_table}. In addition we place upper limits on the remaining pulsars in the sample. For clarity, we redefine the various models listed in Table \ref{braking_index_table} below. In each timing model, we fitted for the pulsar position, spin ($\nu$) and spin-down parameter ($\dot{\nu}$) in addition to
\begin{itemize}
    \item a power-law timing noise and $\ddot{\nu}$ (PL+F2),
    \item a power-law, $\ddot{\nu}$ and low-frequency components longer than the data set (PL+F2+LFC),
    \item a power-law, $\ddot{\nu}$ and proper motion (PL+F2+PM),
    \item a cut-off power law and a $\ddot{\nu}$ (CPL+F2).
\end{itemize}

\begin{figure*}
\centering
\includegraphics[angle=0,width=\textwidth]{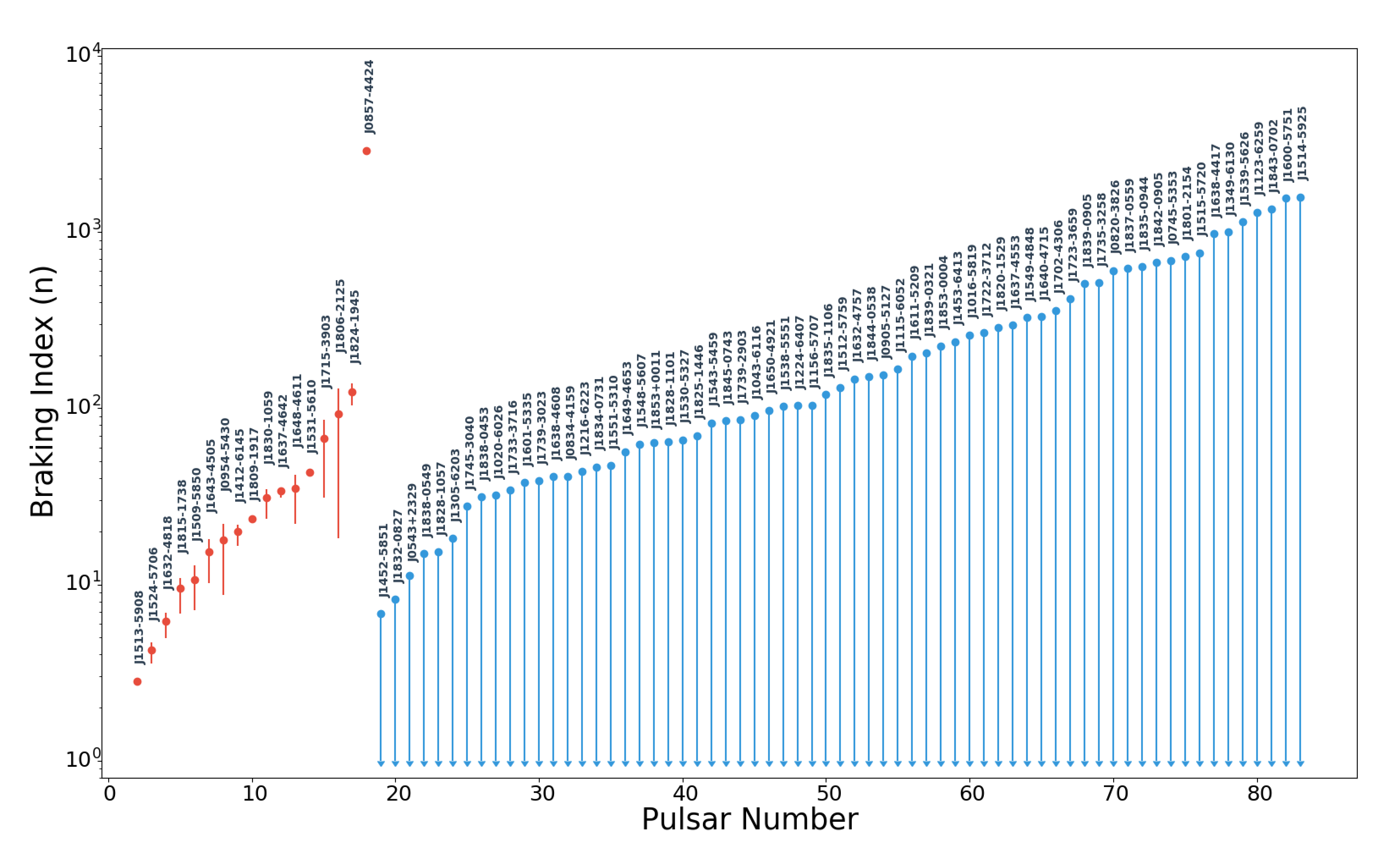}
\caption{\label{brakingindexdist} Braking index detections (red points) and upper limits (blue points) on our sample of 85 pulsars, excluding the 2 pulsars with negative braking indices. The error bars are 97.5\% and 2.5\% limits on the detections and 97.5\% upper limits on the non-detections.}
\end{figure*}

\noindent Figure \ref{brakingindexdist} shows the derived values of $n$ (Equation \ref{braking_index}) and the 97.5\% upper limits.

\begin{table*}
\caption{\label{braking_index_table} The timing noise parameters $A_{\rm red}$, $\beta$ (as defined in equation \ref{red_noise}), the measured spin parameters ($\nu$, $\dot{\nu}$, $\ddot{\nu}$) along with the braking index ($n$) values and the preferred model (as described in Section \ref{measured_BI}) for 19 pulsars.}
      
\centering
\renewcommand{\arraystretch}{1.5}
\resizebox{\textwidth}{!}{
\begin{tabular}{lrrrrrrrrrrr}
\hline
\hline
PSR & $\log_{\rm 10}(A_{\rm red})$ & $\beta$ & $\nu$ & $\dot{\nu}$ & $\ddot{\nu}$ & $n$ & Preferred Model \\
 & (yr$^{3/2}$) & & $(s^{-1})$ & $(10^{-14} s^{-2})$ & $(10^{-23}s^{-3})$ &  & \\
\hline
J0857--4424 & ${-11.3}^{+1.2}_{-0.6}$ & ${-9.1}^{+3.8}_{-1.6}$ & 3.0601045423(4) & --19.6145(10) & 3.63(16) & 2890(30) & PL+F2 \\
J0954--5430 & ${-10.4}^{+0.6}_{-0.3}$ & ${-4.4}^{+2.1}_{-0.8}$ & 2.11483307064(18) & --19.6358(5) & 0.032(8) & 18(9) & PL+F2 \\
J1412--6145 & ${-10.7}^{+1.1}_{-0.6}$ & ${-7.9}^{+3.6}_{-1.6}$ & 3.1720007909(13) & -99.643(4) & 0.62(4) & 20(3) & PL+F2 \\
J1509--5850 & ${-11.1}^{+3.1}_{-2.1}$ & ${-5.1}^{+8.6}_{-2.9}$ & 11.2454488757(7) & --115.9175(16) & 0.12(16) & 11(3) & PL+F2 \\
J1513--5908 & ${-9.7}^{+0.4}_{-0.2}$ & ${-5.7}^{+1.3}_{-0.6}$ & 6.59709182778(19) & --6653.10558(27) & 189.6(2) & 2.82(6) & PL+F2 \\
J1524--5706 & ${-10.2}^{+1.0}_{-0.7}$ & ${-3.6}^{+3.6}_{-1.3}$ & 0.89591729463(9) & --28.60366(2) & 0.038(2) & 4.2(7) & PL+F2 \\
J1531--5610 & ${-11.8}^{+1.3}_{-0.6}$ & ${-8.5}^{+3.4}_{-1.6}$ & 11.8756292823(4) & --194.5360(14) & 1.37(2) & 43(1) & PL+F2 \\
J1632--4818 & ${-9.6}^{+0.8}_{-0.5}$ & ${-5.0}^{+2.7}_{-1.1}$ & 1.2289964712(14) & --98.0730(3) & 0.48(4) & 6(1) & PL+F2 \\
J1637--4642 & ${-9.7}^{+0.6}_{-0.3}$ & ${-4.9}^{+2.2}_{-0.9}$ & 6.491542203(4) & --249.892(10) & 3.2(15) & 34(3) & PL+F2 \\
J1643--4505 & ${-10.1}^{+0.5}_{-0.3}$ & ${-2.3}^{+1.0}_{-0.4}$ & 4.212470392(4) & --56.473(10) & 0.11(2) & 15(6) & PL+F2+LFC \\
J1648--4611 & ${-10.4}^{+0.8}_{-0.5}$ & ${-6.3}^{+2.3}_{-0.9}$ & 6.0621606076(2) & --87.220(5) & 0.44(8) & 40(10) & PL+F2 \\
J1715--3903 & ${-9.2}^{+0.2}_{-0.1}$ & ${-3.8}^{+1.3}_{-0.6}$ & 3.5907423095(9) & --48.2784(13) & 0.4(11) & 70(40) & PL+F2 \\
J1738--2955 & ${-9.6}^{+0.5}_{-0.2}$ & ${-5.8}^{+2.4}_{-1.0}$ & 2.2551713364(2) & --41.7146(12) & --0.51(16) & --70(40) & PL+F2 \\
J1806--2125 & ${-9.1}^{+0.3}_{-0.1}$ & ${-6.6}^{+1.6}_{-0.7}$ & 2.075444041(15) & --50.821(2) & 1.1(4) & 90(60) & PL+F2\\
J1809--1917 & ${-11.7}^{+1.1}_{-0.6}$ & ${-9.0}^{+3.5}_{-1.4}$ & 12.0838226201(8) & --372.7882(19) & 2.70(3) & 23.5(6) & PL+PM+F2 \\
J1815--1738 & ${-11.8}^{+3.1}_{-1.5}$ & ${-4.5}^{+3.1}_{-1.4}$ & 5.03887545888(10) & --197.4552(11) & 0.73(8) & 9(3) & PL+F2+LFC \\
J1824--1945 & ${-10.9}^{+0.3}_{-0.1}$ & ${-3.4}^{+0.6}_{-0.3}$ & 5.281575552287(3) & --14.6048(5) & 0.05(2) & 120(20) & PL+F2+LFC \\
J1830--1059 & ${-8.5}^{+0.3}_{-0.1}$ & ${-13.6}^{+6.2}_{-2.8}$ & 2.4686900068(5) & --36.5201(10) & 0.167(19) & 31(7) & CPL+F2 \\
J1833--0827 & ${-10.2}^{+0.2}_{-0.1}$ & ${-2.8}^{+1.2}_{-0.6}$ & 11.7249580817(4) & --126.1600(8) & --0.198(13) & --15(2) & PL+F2 \\
\hline
\end{tabular}}
\end{table*}

Of the 19 braking index measurements, four of them have $n<10$, eleven have $10<n<100$, two of them are $>$ 100 and two of them are negative. Only for PSR~J1513--5908 is the value close to $n=3$ as expected from a simple model of a magnetic dipole in vacuum. Two questions thus arise. The first is whether the measured values of $n$ are robust to the passage of time and the second is whether or not unseen glitches affect the measured values of $n$? We answer these questions in the next three sections.

\section{Robustness of the braking index measurements} \label{robustness_BI}
\subsection{Comparisons with simulated data sets.}
We first test the efficacy of \textsc{temponest} (\citealt{temponest}) in this context, by simulating timing residuals for the 19 pulsars discussed in this paper with various induced signatures as per the preferred model and check the consistency with which we recover the input parameters. The simulated timing residuals are generated using \textsc{libstempo}\footnote{\url{https://github.com/vallis/libstempo}} with data spans similar to the actual observations. The times of arrival (ToAs) are considered to be a combination of deterministic and stochastic components. White noise components are modelled by adjusting the uncertainty on a ToA to be, 
  \begin{equation} \label{white_noise}
  \sigma^{2} = F{\sigma_{\rm r}}^{2} + {\sigma_{\rm Q}}^{2},
  \end{equation}
where the parameter $F$, referred to as the EFAC, accounts for instrumental distortions and ${\sigma_{\rm r}}^{2}$ is the formal uncertainty obtained from ToA fitting. An EQUAD ($\sigma_{\rm Q}^{2}$) is introduced to model time independent noise per observing system. 
The timing noise is modelled to be a power-law described by a red-noise amplitude ($A_{\rm red}$) and spectral index ($\beta$),
  \begin{equation} \label{red_noise}
  P_{\rm r}(f) = \frac{A_{\rm red}^2}{12\pi^2} \left( \frac{f}{f_{\rm yr}}\right )^{\rm -\beta} ,
  \end{equation}
where $f_{yr}$ is a reference frequency of 1 cycle per year and $A_{\rm red}$ is in units of yr$^{3/2}$. A model with a cut-off power law was favoured for PSR J1830--1059 which was simulated following the power spectrum,
  \begin{equation} \label{cutoff-pl}
  P_{\rm r,CF}(f) = \frac {A (f_{\rm c}/f_{\rm yr})^{\rm -\beta}}{[1+(f/f_{\rm c})^{\rm -\beta/2}]^2},
  \end{equation}
where $f_{\rm c}$ is the corner frequency and $A$ is $(A_{\rm red}^2/12\pi^2)$.

To maintain consistency in the simulations, a model with low-frequency components (LFC) with timescales longer than the data set, were simulated for PSRs J1643--4505, J1815--1738 and J1824--1945 and a model with a proper motion signature was simulated for PSR J1809--1917. Finally, the simulated timing residuals were modified to include a braking index signature.

\begin{figure*}
\centering
\includegraphics[angle=0,width=1\textwidth]{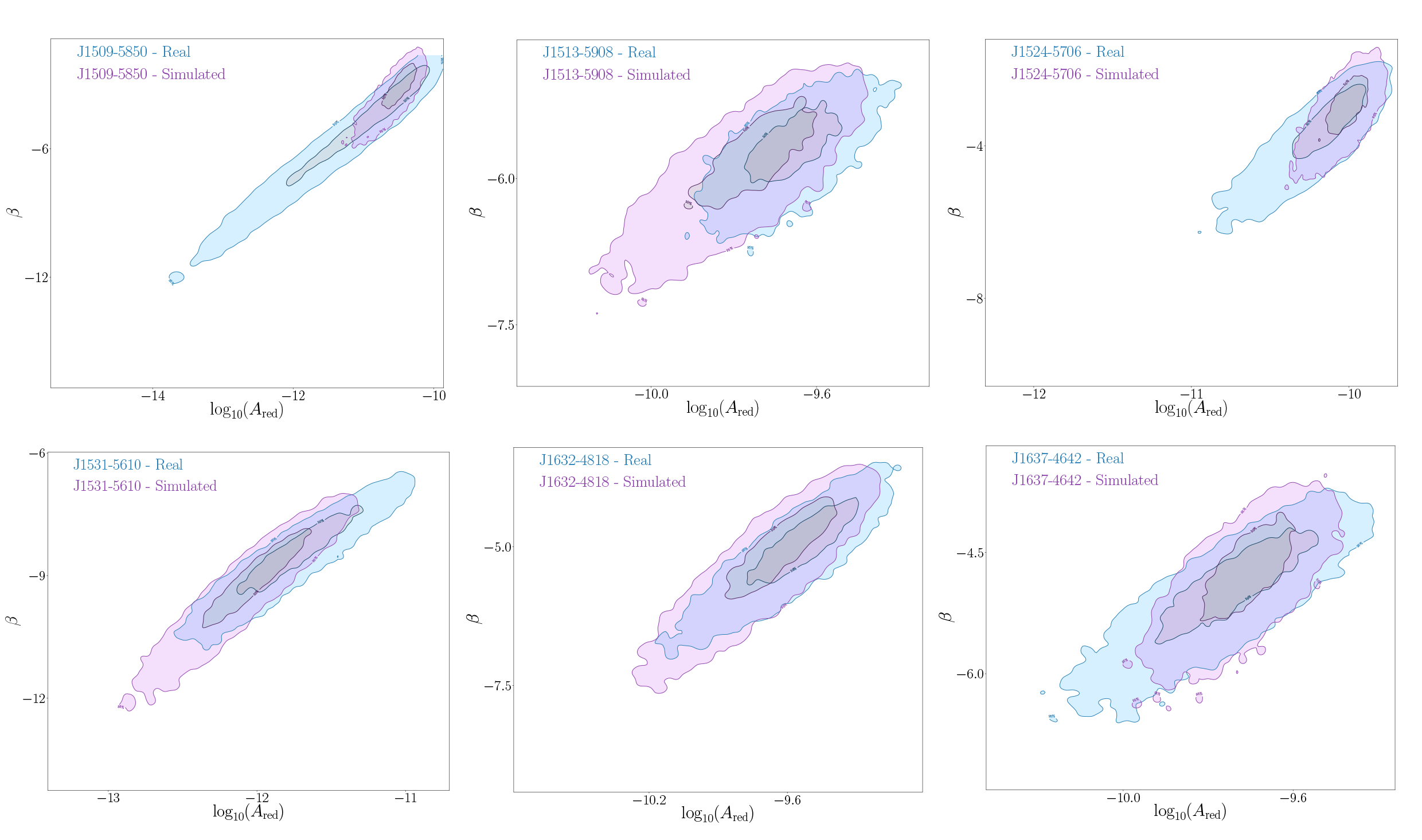}
\caption{\label{ampslope_rvs} Posterior distributions of the red noise amplitudes and spectral indices for a sample of six pulsars. The blue posteriors represent real data, while the purple posteriors represent simulated data.}
\end{figure*}

\begin{figure*}
\centering
\includegraphics[angle=0,width=1\textwidth]{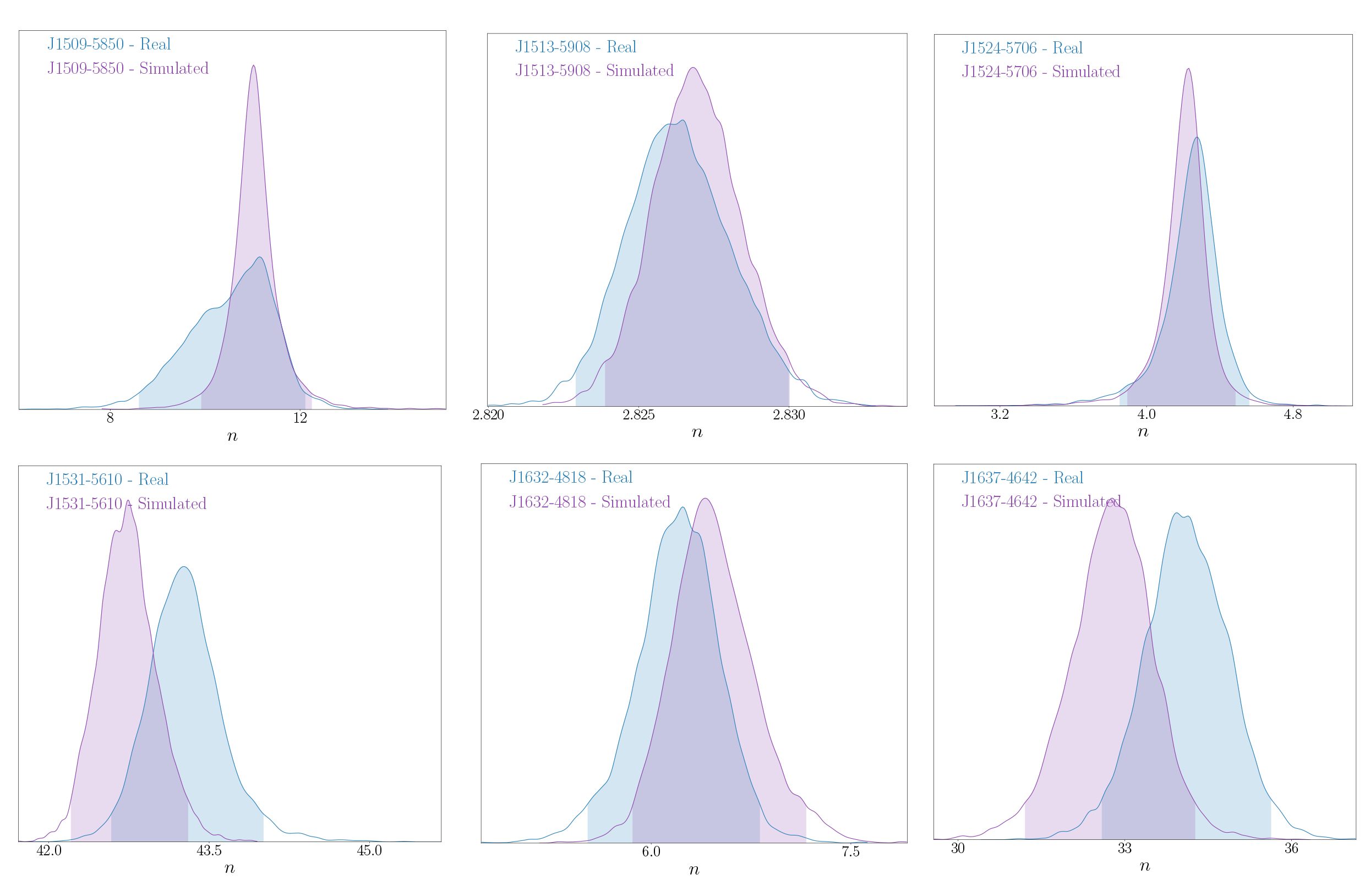}
\caption{\label{f2brake_rvs} Posterior distributions of the braking index measurements for a sample of six pulsars. The blue posteriors represent real data, while the purple posteriors represent simulated data.}
\end{figure*}

We use the Bayesian pulsar timing package, \textsc{temponest} to search for the stochastic parameters and the braking index value in the simulated data sets. 
The prior ranges on the stochastic parameters, corner frequency, low-frequency components, $\ddot{\nu}$ and the proper motion are reported in Table \ref{tab:prior_ranges}.
Our measurements of the stochastic, braking index and other induced parameters (proper motion, corner frequency and LFC) are consistent between the observed and simulated data sets for all 19 pulsars. Figures \ref{ampslope_rvs} and \ref{f2brake_rvs} show the consistency in the measured values of red noise amplitudes, spectral indices, and the derived braking index values from the real and simulated data sets for a sample of six pulsars. The shaded regions in the plots represent 50\% and 95\% confidence intervals. 

\begin{table}
\caption{\label{tab:prior_ranges}
Prior ranges for the various stochastic and deterministic parameters used in the timing models. $\Delta_{\rm param}$ is the uncertainty on the marginalized timing model parameters from the initial \textsc{tempo2} fitting.}
\centering
\renewcommand{\arraystretch}{1.5}
\resizebox{\columnwidth}{!}{
\begin{tabular}{lrrrrrrrrrrr}
\hline
\hline
Parameter & Prior range & Type  \\
& &  \\
\hline
Red noise amplitude ($A_{\rm red}$) & (-20,-5) & Log-uniform \\ 
Red noise slope ($\beta$) & (0,20) & Log-uniform \\
EFAC & (-1,1.2) & Log-uniform \\
EQUAD & (-10,-3) & Log-uniform \\ 
Corner frequency ($f_{\rm c}$) & (0.01/$T_{\rm span}$,10/$T_{\rm span}$) & Log-uniform \\
Low frequency cut-off (LFC) & (-1,0) & Log-uniform \\
$\ddot{\nu}$ & $\pm$ 10000 $\times\Delta_{\rm param}$ & Uniform \\ 
Proper motion & $\pm$ 1000 mas/yr & Uniform \\ 
\hline
\end{tabular}}
\end{table}

\subsection{Using legacy data sets.} \label{legacy_section}
\begin{figure}
\centering
\includegraphics[angle=0,width=0.45\textwidth]{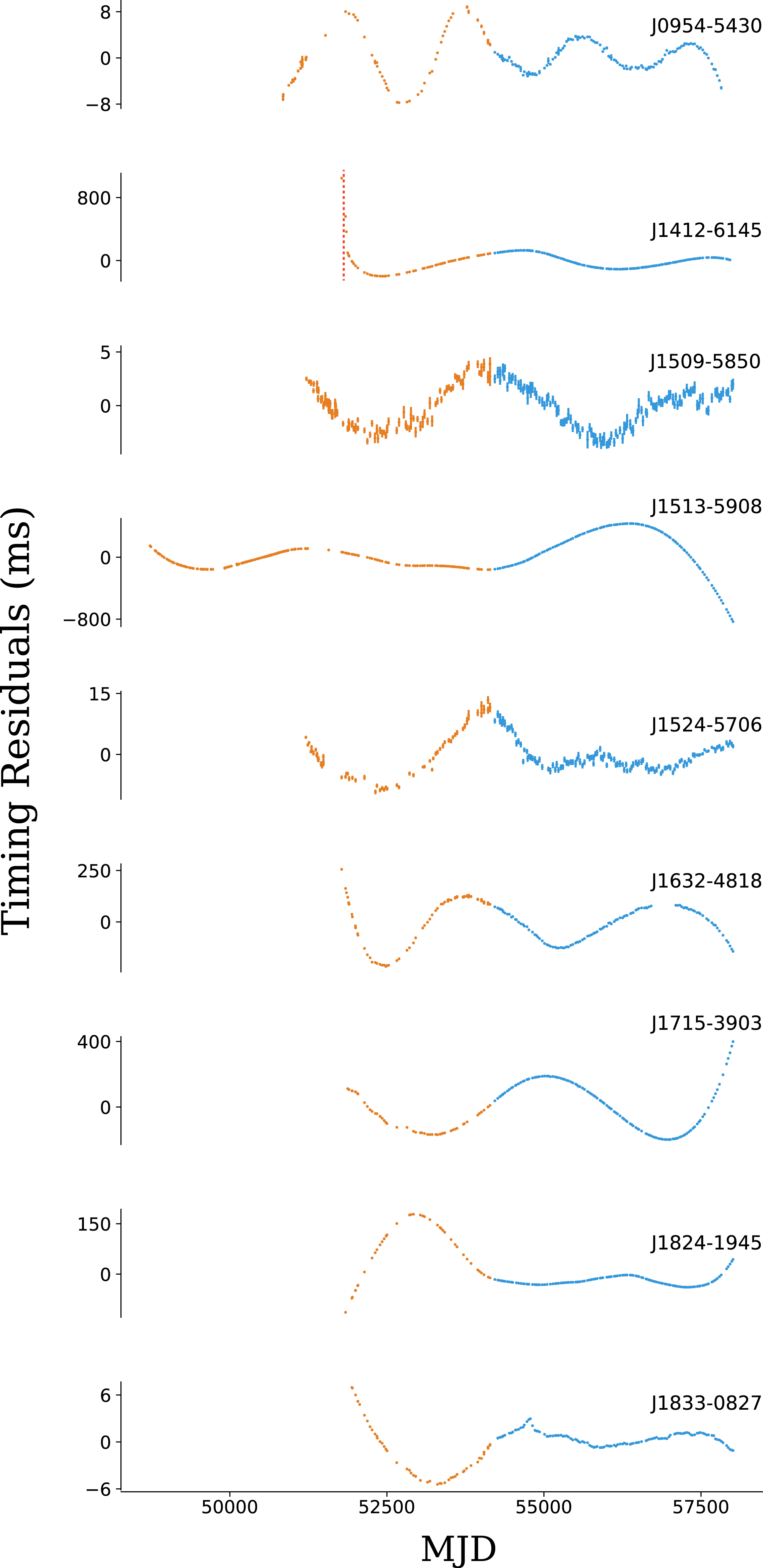}
\caption{\label{timingplots_10psrs} Phase connected timing residuals for nine pulsars with added archival data sets (orange) are shown here before subtracting the timing noise model. The vertical dashed line in PSR J1412--6145 shows the epoch of the glitch reported in \protect\cite{yu_glitches}. For PSRs J1632--4818 and J1715--3903, the new glitches described in Section \ref{legacy_section} are modelled in these timing residual plots.}
\end{figure}

We obtained historical data for 9 of the 19 pulsars and prepended these data to those used in PSJ19. The historical data is also at 20-cm wavelengths observed with the 64-m CSIRO Parkes radio telescope. The data set used in PSJ19 is hereafter called the PSJ19-data set and the data set that includes the historical data is called the legacy data set. Figure~\ref{timingplots_10psrs} shows the timing residuals for these nine pulsars with the added data. 

\begin{figure*}
\centering
\includegraphics[angle=0,width=1\textwidth]{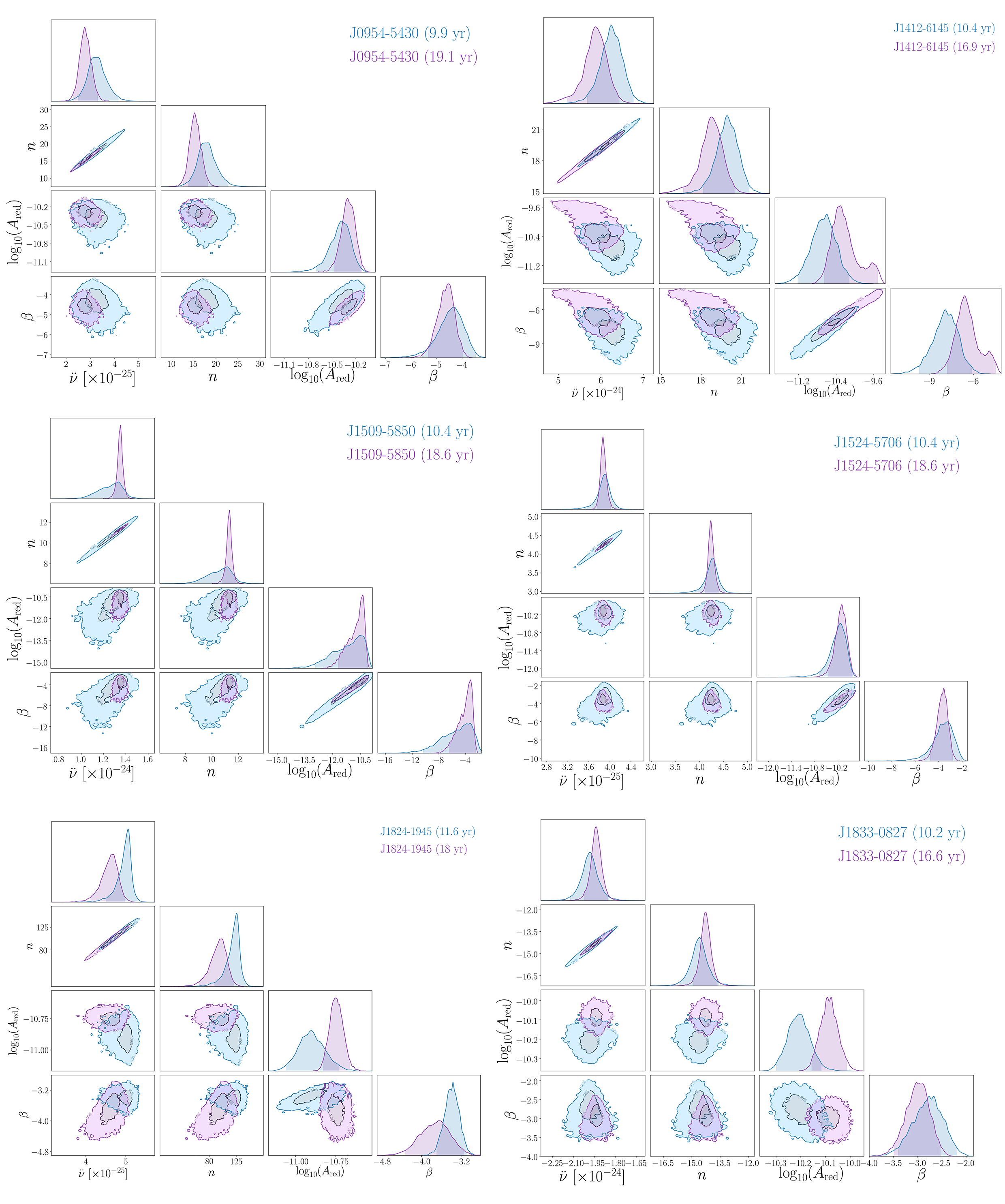}
\caption{\label{consistent_history} Posterior distributions of six pulsars comparing the original (in blue) and the longer (legacy) data sets (in purple). We account for a previously published glitch at MJD $\sim$ 51868 for  PSR J1412--6145.}
\end{figure*}

Figure \ref{consistent_history} shows the posterior distributions for six pulsars that have consistent values (overlapping confidence intervals) of $\ddot{\nu}$, braking index and the timing noise parameters using the legacy data set. A model with power-law timing noise and $\ddot{\nu}$ (PL+F2) was still favoured for three pulsars and PL+F2+LFC for PSR J1824-1945. For PSR J1412$-$6145, we fitted for the glitch parameters as reported in \cite{yu_glitches} while simultaneously modelling the stochastic and braking index parameters. The preferred model in this case is one with the glitch parameters (PL+F2+GL). For PSR J1833--0827, we measure a proper motion signal with the legacy data set, with the proper motion in right ascension ($\mu_{\alpha}$) being $-37\pm$7 mas yr$^{\rm -1}$, in declination ($\mu_{\delta}$) being 3$\pm$26 mas yr$^{\rm -1}$ and the total proper motion ($\mu_{\rm tot}$) being $39\pm10$~mas~yr$^{\rm -1}$. At a distance of 4.38 kpc (from DM estimates), this amounts to a high transverse velocity of 800$\pm$200~kms$^{\rm -1}$. The preferred timing model with the legacy data set is PL+F2+PM with a log Bayes factor of 21. It must be noted that the measured braking index remains unchanged within uncertainties.
PSRs J1513--5908, J1632--4818 and J1715--3903 exhibited clear inconsistencies in either the $\ddot{\nu}$ or the timing noise parameters after adding archival data. These are further investigated below on a case-by-case basis.

{\bf PSR J1513--5908:}
This pulsar has not been observed to glitch. With 28.4 years of radio timing data from 1982 June to 2010 November, \cite{Livingstone_kaspi_2011} reported a braking index value of $2.832 \pm 0.003$. They also reported a detection of the third derivative of the spin-frequency ($\dddot{\nu}$) to be $-0.9139(2)\times10^{-31}$~s$^{-4}$. 

We notice that a model with a $\dddot{\nu}$ is preferred over the PL+F2 model in the legacy data and including it as part of the timing model results in consistent measurements of $\ddot{\nu}$ and the timing noise parameters. An expression for $\dddot{\nu}$ can be derived by taking two derivatives of equation \ref{pulsar_spindown} and is written as,
\begin{equation}
\dddot{\nu}=\frac{n(2 n-1) \dot{\nu}^{3}}{\nu^{2}}.
\end{equation}
The validity of the spin-down law can then be tested by calculating the expected second braking index ($m_{\rm 0}$), which is written as,
\begin{equation}
m_{0} \equiv n(2 n-1),
\end{equation}
and if equation \ref{pulsar_spindown} accurately describes the spin-down of the pulsar, then $m_{0} = m$, where $m$ is,
\begin{equation}
m=\frac{\nu^{2} \dddot{\nu}}{\dot{\nu}^{3}}.
\end{equation}

\begin{figure*}
\centering
\includegraphics[angle=0,width=1\textwidth]{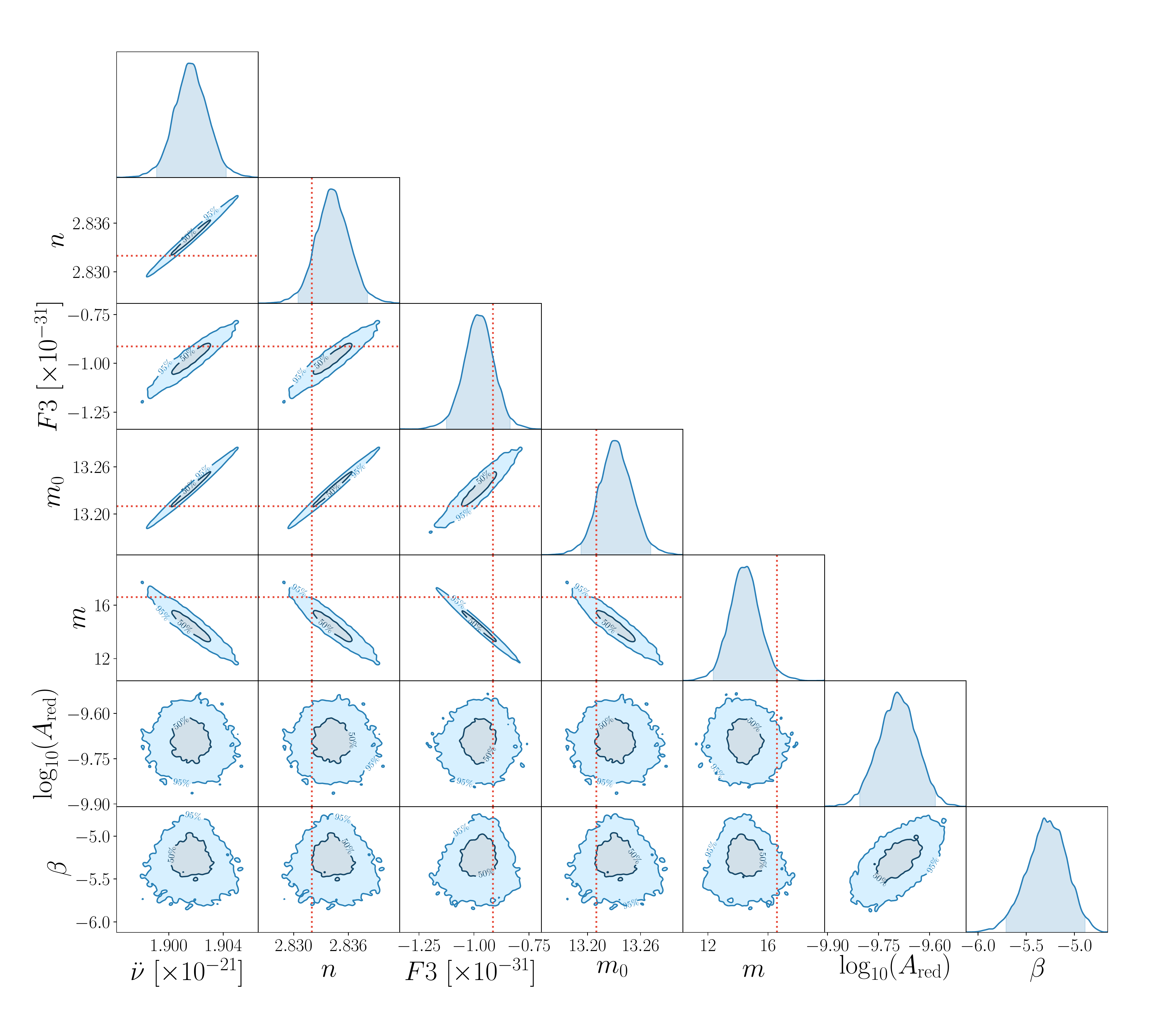}
\caption{\label{1513_withf3}  Posterior distributions of $\ddot{\nu}$, $\dddot{\nu}$, $n$, $m_{0}$, $m$ and the timing noise parameters for PSR J1513--5908 measured from 25.4 years of radio timing data. The dashed orange lines are measurements from \protect\cite{Livingstone_kaspi_2011}.}
\end{figure*}

A value of $m$ can be measured from higher order frequency derivatives of a pulsar, which is only possible in the youngest pulsars that have large spin-down and minimal timing noise. We measure the value of $m$ to be $14.4 \pm 2.2$ and the computed value of $m_{0}$ to be $13.23 \pm 0.03$, which are in agreement with each other within uncertainties. In Figure \ref{1513_withf3} we show our measurements of $\ddot{\nu}$, $\dddot{\nu}$ (F3), $n$, $m_{0}$, $m$ and the timing noise parameters which are consistent with the measurements reported in \cite{Livingstone_1513} within 95\% confidence limits. Our measurements of timing noise and the absence of glitches in this pulsar are in contrast to the suggestions made by \cite{hobbs_2010} that timing noise in pulsars with characteristic ages less than 10$^{5}$ yr is due to unmodelled glitch recovery. 

\begin{figure*}
\centering
\includegraphics[angle=0,width=1\textwidth]{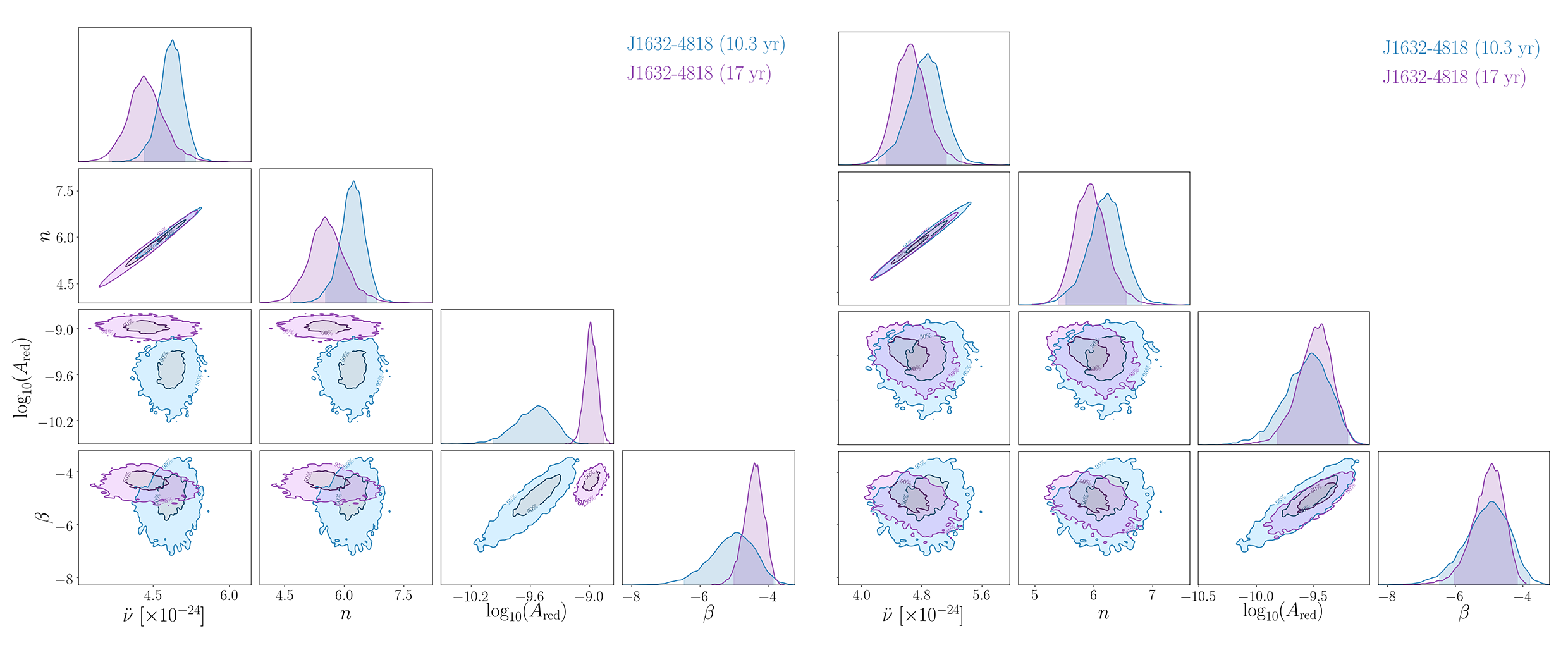}
\caption{\label{1632_consistent} Posterior distributions for PSR J1632--4818, for the PSJ19 (in blue) and legacy data sets (in purple) before (left) and after (right) modelling the glitch.}
\end{figure*}

{\bf PSR J1632--4818:}
Using \textsc{temponest}, we searched for glitches in the legacy data set and found evidence for the presence of a glitch at MJD 53962$\pm$10 days and with a glitch amplitude of (1.1$\pm$0.2$)\times$10$^{-8}$ Hz. The detection was also supported by the fact that the log Bayes factor for the glitch-search model was 17, which is much higher than our Bayes factor threshold of 5 (used in PSJ19). Figure \ref{1632_consistent} shows the posterior distributions of timing noise, $\ddot{\nu}$ and braking index measurements for the PSJ19 and legacy data sets, before and after accounting for the glitch. 

\begin{figure*}
\centering
\includegraphics[angle=0,width=1\textwidth]{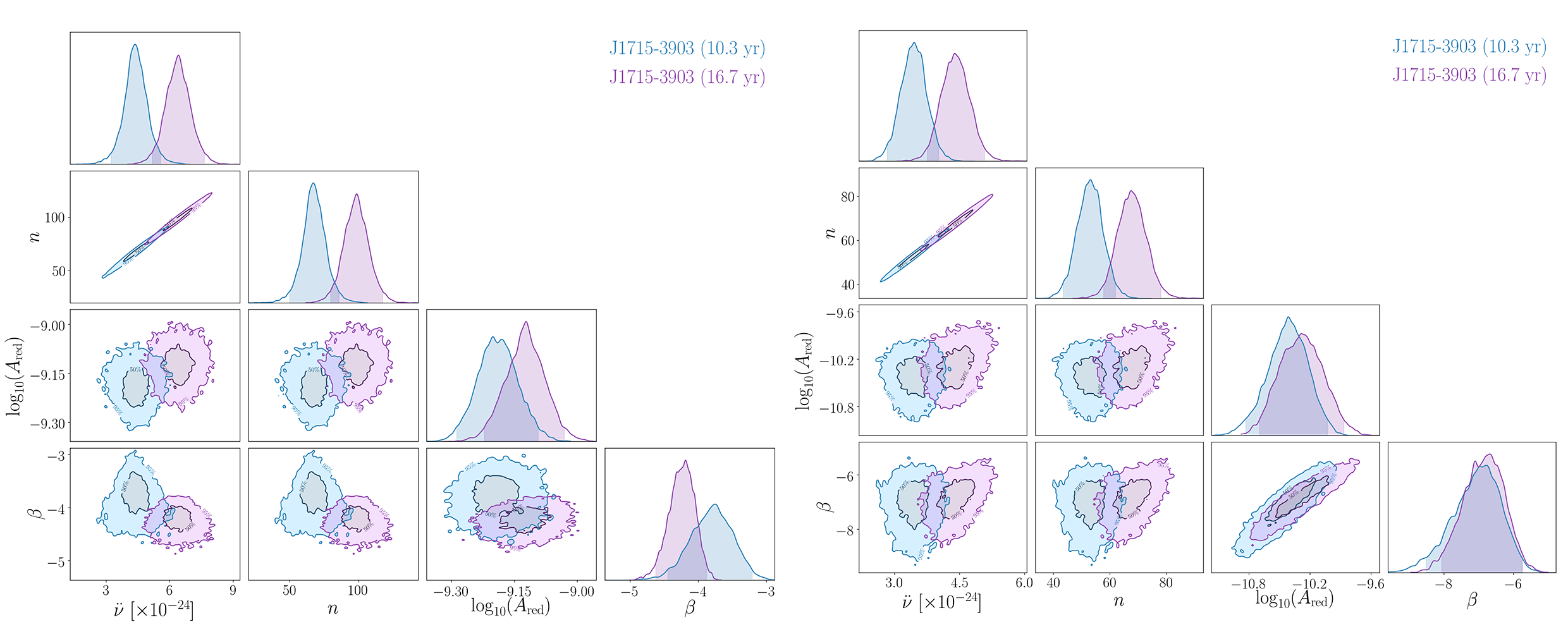}
\caption{\label{1715_consistent} Posterior distributions for PSR J1715--3903, for the PSJ19 (in blue) and legacy (in purple) data sets before (left) and after (right) modelling the glitch.}
\end{figure*}

{\bf PSR J1715--3903:}
Similar to PSR~J1632--4818, we found evidence for the presence of a glitch at MJD 57724$\pm$2 days in the PSJ19 data set, with a glitch amplitude of (1.9$\pm$0.1$)\times$10$^{-8}$ Hz and the glitch model is strongly preferred over a model without the glitch with a Bayes factor of 35. Figure \ref{1715_consistent} shows the measured posterior distributions of $\ddot{\nu}$, braking index and the timing noise parameters for both the data sets before and after modelling for the glitch. It is evident from Figure \ref{1715_consistent} that the uncertainties on the measurement of the braking index have improved after accounting for the glitch. 

\subsection{Summary}
In Table \ref{tab:braking_robust}, we report the time-span of the legacy data sets along with the updated braking index values using the updated timing models for PSR J1412--6145, J1513--5908, J1632--4818 and J1715--3903. 
There are three important results from this section. First, we showed through simulations that {\sc temponest} is capable of recovering braking indices in the presence of strong timing noise. Secondly, we demonstrated that {\sc temponest} is capable of detecting glitches within the time span of the datasets. Thirdly, we demonstrated that increasing the overall data span maintains the value of $n$ even in the presence of glitch activity.

\begin{table*}
\caption{\label{tab:braking_robust}
Braking index measurements for nine pulsars before and after increasing the time span of the data as measured from the preferred timing models.}. 
\centering
\resizebox{\textwidth}{!}{
\begin{tabular}{lrrrrrrrrrrr}
\hline
\hline
PSR & Old timespan & New MJD Range & New timespan & Old braking index & New braking index & Old timing model & New timing model \\
& (yr) & & (yr) & & & & \\
\hline
J0954--5430 & 10 & 50849--57824 & 19 & 18(9) & 15(3) & PL+F2 & PL+F2 \\
J1412--6145 & 10 & 51782--58012 & 16.9 & 20(3) & 19(2) & PL+F2 & PL+F2+GL \\
J1509--5850 & 10 & 51221--58012 & 18.5 & 11(3) & 11.2(5) & PL+F2 & PL+F2\\ 
J1513--5908 & 11.6 & 48732--58012 & 25 & 2.826(3) & 2.834(3) & PL+F2 & PL+F2+F3\\
J1524--5706 & 10 & 51212--58012 & 18.6 & 4.3(7) & 4.2(1) & PL+F2 & PL+F2 \\
J1632--4818 & 10 & 51782--58012 & 14 & 6.2(7) & 5.9(5) & PL+F2 & PL+F2+GL \\ 
J1715--3903 & 10 & 51878--58012 & 14 & 70(40) & 68(10) & PL+F2 & PL+F2+GL  \\ 
J1824--1945 & 11.6 & 51844--58012 & 18 & 120(20) & 88(15) & PL+F2+LFC & PL+F2+LFC \\
J1833--0827 & 10 & 51782--58012 & 16.6 & $-$15(2) & $-$14.2(6) & PL+F2 & PL+F2+PM \\
\hline
\end{tabular}}
\end{table*}

\section{Glitch recovery} \label{glitch_recovery_BI}
Here we investigate whether unseen glitches in the past affect the measured value of $n$. Of crucial importance is not the glitch size itself but rather the recovery from the glitch. Generally the form of the recovery is taken to be an exponential which asymptotes towards a frequency increment parameter, $\nu_{\rm GLD}$, with a glitch decay time of $\tau_{\rm GLD}$. Typically, glitch recovery is expressed as $Q \equiv \nu_{\rm GLD}/\nu_{\rm g}$, where $\nu_{\rm g}$ is the total increment in the pulsar spin-frequency due to the glitch, expressed as $\nu_{\rm GLD} + \nu_{\rm p}$. Here, $\nu_{\rm p}$ is the permanent increment in the pulse frequency. The exponential form of the recovery implies that the epoch of the glitch is not important and since we are interested in glitches before our data set, the $\nu_{\rm p}$ is set to zero. In what follows therefore, the glitch is assumed to occur one day prior to the start of our data sets. We also keep in mind that the presence of $\ddot{\nu}$ (and hence a braking index) manifests itself as a cubic in the timing residuals.

We can imagine several cases of interest. First, it is clear that very small values of $\nu_{\rm GLD}$ and/or small decay times will have little effect on our ability to measure $\ddot{\nu}$. For large decay times (longer than our dataset) the exponential looks linear and will be absorbed into the fits of $\nu$ and $\dot{\nu}$ and will hence not affect $\ddot{\nu}$. However, for decay times comparable to our data span (i.e. from 1000 days to 5000 days) the exponential mimics a (partial) cubic, and can therefore masquerade as a $\ddot{\nu}$ term. This will always serve to {\bf increase} $\ddot{\nu}$ and hence also increase the value of $n$. Larger values of $\nu_{\rm GLD}$ will lead to larger values of $n$. In general, the induced value of $n$ is largely independent of the other pulsar parameters; it depends only on $\nu_{\rm GLD}$ and $\tau_{\rm GLD}$. However, the presence of timing noise complicates such analysis. 

We assume the presence of an unseen glitch before the data set for each of our 19 pulsars and search for evidence of glitch recovery signals spanning a $\nu_{\rm GLD}$ range of $\mathrm{10^{-14}}$ to $\mathrm{10^{-3}}$ Hz and a $\tau_{\rm GLD}$ range of 1 to 10,000 days. We use a log-uniform prior for the glitch decay parameters. The prior ranges for the timing noise parameters and $\ddot{\nu}$ are as reported in Table \ref{tab:prior_ranges}. We then fitted for three different timing models and compared their log-evidences. These are: 
\begin{itemize}
    \item the timing model as reported in Tables \ref{braking_index_table} and \ref{tab:braking_robust},
    \item the timing model with a glitch recovery signal (TimingModel+GLR)
    \item the timing model, without a cubic polynomial (F2), but including a glitch recovery signal (TimingModel-F2+GLR)
\end{itemize}

For PSR J0954--5430, the three models that we compare are PL+F2, PL+F2+GLR and PL+GLR, since the timing model as reported in Table \ref{braking_index_table} is PL+F2. However, for PSR J1412--6145, since the updated timing model reported in Table \ref{tab:braking_robust} is PL+F2+GL, the three models to compare are PL+F2+GL, PL+F2+GL+GLR and PL+GL+GLR (without the F2). 

Table~\ref{glitch_recovery_table} reports the log-evidences for the three timing models for 19 pulsars in our sample. The preferred model for each pulsar is highlighted. Two things are apparent; first that there is much stronger evidence for a cubic term than for an exponential. Secondly, that although TimingModel+GLR and TimingModel have similar log-evidences, the simpler model would always be preferred. This is following the Occam principle, which states that all other things being equal, a simpler model is more likely than a complicated one. Taken together, this shows that there is little evidence for an exponential signal in the data. It is also important to note that the Bayesian inference framework enables us to identify cases when we cannot distinguish between models. In this case, distinguishing between a braking index and a glitch recovery signal is important while interpreting the measurements. 

Finally we note that PSRs J1531--5610, J1806--2125, J1809--1917 and J1833--0827 have glitches detected before our data span. The glitch sizes are a few $\times 10^{-5}$~Hz as reported in \cite{yu_glitches} and \cite{espinoza_305} and the difference between the MJD of the first ToA in our data set and the glitch epochs are $\sim$ 2500, 2300, 1000 and 4200 days respectively. \cite{yu_glitches} also measure glitch recovery times of a few hundred days for 3 of these pulsars. If correct, then no evidence of the glitch would remain in the residuals of our data as is confirmed by Table~\ref{glitch_recovery_table}.

\begin{table}
\caption{\label{glitch_recovery_table}
Comparison of log-evidence values for models with glitch recovery and braking index for 19 pulsars. The preferred model for each pulsar is highlighted.} 
\resizebox{\columnwidth}{!}{
\begin{tabular}{lrrrrrrrrrrr}
\hline
\hline
PSR & TimingModel+GLR & TimingModel & TimingModel-F2+GLR  \\
& (with F2 \& GLR) & (with only F2) & (with only GLR) \\
\hline
J0857--4424 & 1144  & \textbf{1150} & 807 \\
J0954--5430 & 1357 &\textbf{1358} & 934 \\
J1412--6145 & 1184 &\textbf{1197} & 1165 \\
J1509--5850 & 1593 &\textbf{1594} & 1565 \\
J1513--5908 & 2961 &\textbf{2962} & 2452 \\
J1524--5706 & 1156 &\textbf{1157} & 721 \\
J1531--5610 & 1082 &\textbf{1083} & 1056 \\
J1632--4818 & 717 & \textbf{725} & 460 \\
J1637--4642 & 727 & \textbf{728} & 684 \\
J1643--4505 & 705 & \textbf{706} & 698 \\
J1648--4611 & 739  &\textbf{746}& 707 \\
J1715--3903 & 893 & \textbf{901} & 725 \\
J1738--2955 & 510 & \textbf{511} & 481 \\
J1806--2125 & 572 & \textbf{572} & 528 \\
J1809--1917 & 943  &\textbf{946} & 774 \\
J1815--1738 & 517 & \textbf{518} & 510 \\
J1824--1945 & 1546 &\textbf{1548} & 986 \\
J1830--1059 & 958 & \textbf{960} & 943 \\
J1833--0827 & 1345 &\textbf{1346} & 617 \\
\hline
\end{tabular}}
\end{table}

\section{Glitch simulations} \label{glitch_sim_sec}
We have shown in the above section that there is no (Bayesian) evidence for the presence of an exponential glitch recovery signal in our data, even in those pulsars known to have glitched prior to our data set. However, we also noted that a long decay time ($>1000$ days) could induce a positive $n$ in the residuals. In this section, we therefore consider the possibility of long glitch recovering times contaminating our data.

Using the \textsc{tempo2} package, we measure the induced $\ddot{\nu}$ as determined when fitting for F2 for different simulated data sets with a typical data span of 15~yr, and for 1000 days $< \tau_{\rm GLD} <$ 10000 days. Using these measurements, we empirically determine the link between $\nu_{\rm GLD}$, $\tau_{\rm GLD}$ and the induced $\ddot{\nu}$ as

\begin{equation} \label{induced_f2_glitch}
    \ddot{\nu} = 10^{-14} \,\,\, \frac{\nu_{\rm GLD}}{\tau_{\rm GLD}} s^{-3}
\end{equation}

\noindent with $\nu_{\rm GLD}$ in Hz and $\tau_{\rm GLD}$ in days. 
For a given pulsar this translates into an induced $n$ via Equation~\ref{braking_index}. Second, for each pulsar we compute the glitch rate and typical glitch size. For this we use the metric for the glitch rate given by \cite{fuentes_glitch} along with the parameterisation of large glitch sizes given in the same paper. As a check on the metric, we use Poisson statistics and determine that we expect $\sim$3.2 pulsars in our sample to have glitched in the 15 year observing span of the data. We have shown above that indeed 3 pulsars have undergone glitches in this time span, in good agreement with the predictions of \cite{fuentes_glitch}.

We then perform a Monte Carlo simulation in the following way:
\begin{itemize}
 \item For each pulsar, we determine the glitch rate, and then using Poisson statistics randomly select a date in the past for the glitch to have occurred.
 \item Draw a glitch size from the distribution of large glitches given in \cite{fuentes_glitch}.
 \item For a given $\tau_{\rm GLD}$, extrapolate forwards in time to the start of our data set and then compute the effective glitch size at that date assuming an exponential recovery.
 \item Compute the induced $\ddot{\nu}$ and hence $n$ from Equation \ref{induced_f2_glitch} above.
 \item If the induced $n$ is greater than the upper limit for a given pulsar from Table~\ref{braking_index_table} then count this as a detection.
\end{itemize}
The output of the simulation shows that the most likely `detections' of $n$ are, unsurprisingly, in pulsars with a high glitch rate and/or for which the upper limit on $n$ is small. For example, PSR~J1809--1917 has the highest detection probability of 0.98 with a median induced $n$ of 21 followed by PSR~J1531--5610 with a detection probability of 0.86 and a median induced $n$ of 37.2. Both these pulsars feature in Table~$\ref{braking_index_table}$. The question now arises as to how the addition of timing noise affects these results and whether \textsc{temponest} can detect long timescale glitch recovery signals.

We consider the cases of PSR~J1809--1917 and J1531--5610. We simulate a 10~yr data set with an induced $\nu_{\rm GLD}$ of $4\times10^{-6}$~Hz, a $\tau_{\rm GLD}$ of 5000 days and the glitch epoch set one day before the first data set. A power-law timing noise signal is induced with $A_{\rm red}$ and $\beta$ as in Table~\ref{braking_index_table}. We set $\ddot{\nu}=0$. Fitting for the three different models; PL+F2, PL+GLR and PL+F2+GLR using \textsc{temponest}, we find that in both the cases, the PL+GLR model is preferred over the PL+F2+GLR model (with Bayes factors of 12 and 7 respectively), while the PL+F2 model is strongly disfavoured (with Bayes factors of $\sim$ -200). 

These simulations together with the results from Table \ref{glitch_recovery_table} have shown that:
\begin{itemize}
     \item the braking indices are robust to the addition of legacy data, which is crucial in indicating a lack of exponential signal in the PSJ19 data,
    \item the model with $\ddot{\nu}$ is preferred over an exponential glitch recovery model for all 19 pulsars,
    \item given the glitch size and rate distributions as in \cite{fuentes_glitch},  there are not enough unseen glitches prior to the beginning of our dataset to explain the values of $n$ we obtain even for long recovery times,
    \item for long exponential glitch recovery times comparable to our data span, {\sc temponest} will detect and distinguish the presence of a glitch recovery signal from a cubic term.
\end{itemize}
In conclusion, there is strong evidence for a cubic term in the residuals of the 19 pulsars rather than an exponential term and the measured values of braking indices are unlikely due to an exponential glitch recovery.

\section{Implications} \label{implications_sec}
\begin{figure*}
\centering
\includegraphics[angle=0,width=0.95\textwidth]{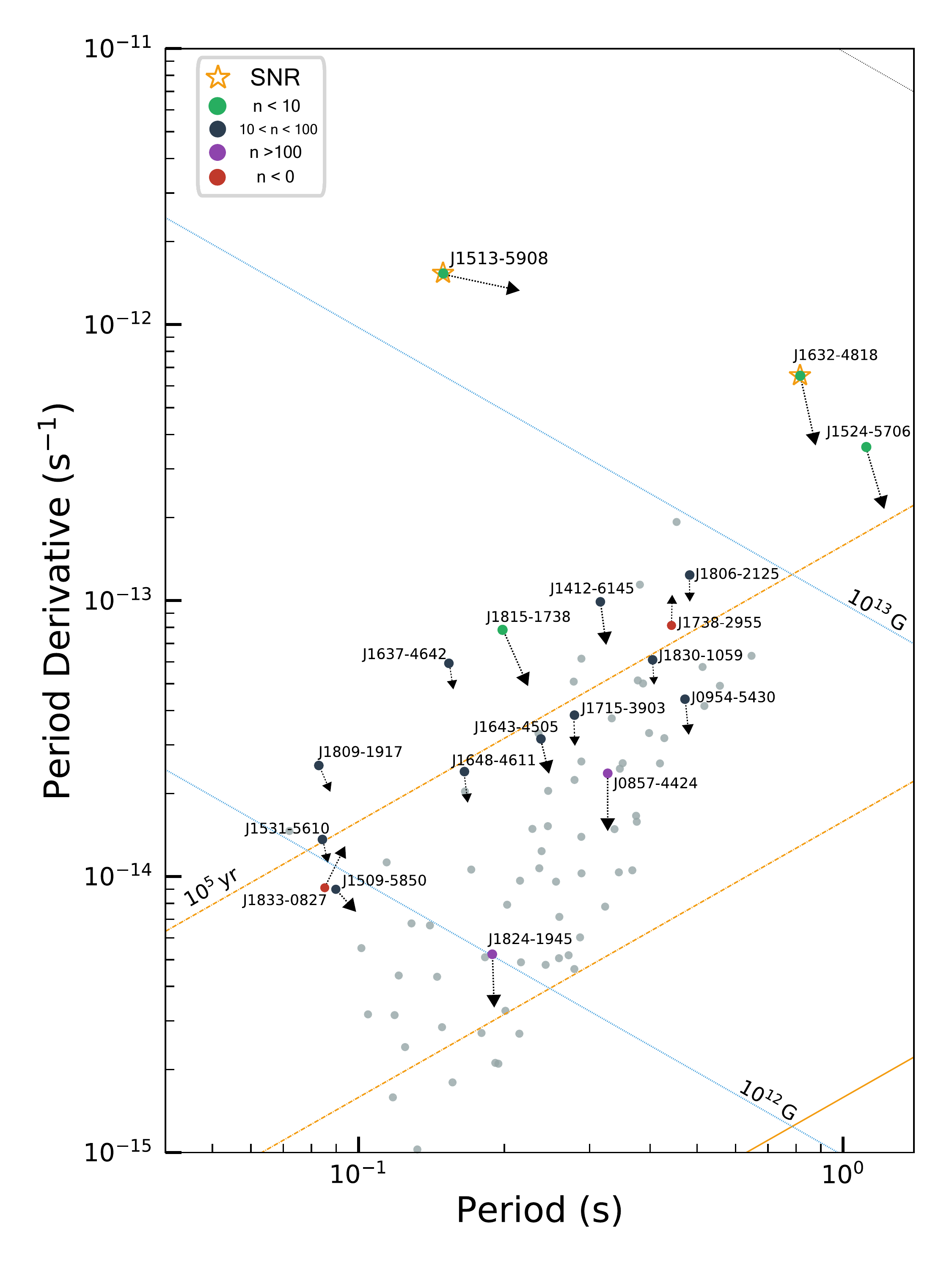}
\caption{\label{ppdot_evolution} The $P$-$\dot{P}$ diagram for the 19 pulsars discussed in this paper (colour-coded filled circles) and the rest of the pulsars discussed in PSJ19 (grey circles). Lines of constant age are in orange (dashed line), while lines of constant magnetic fields are in blue. The dashed arrows show the time-evolution of the pulsars which follow a slope of $2-n$. The length of the arrows are arbitrary to preserve clarity.}
\end{figure*}

In Figure~\ref{ppdot_evolution}, we show evolutionary tracks for each of our pulsars given their measured braking index. From equation \ref{pulsar_spindown}, if $K$ and $n$ are assumed to be constant in time, then the pulsars follow a track with slope $2-n$ in the $P$-$\dot{P}$ diagram. So, for pure magnetic braking with a constant dipolar magnetic field, we expect neutron stars to follow lines with slope of $-1$. However, previously measured values of $n$ have been different from the expected value of 3 and in this paper we have reported $n \gg 3$, implying that neutron stars can exhibit substantial torque decay and that the simple dipolar spin-down model is clearly imperfect. 

\begin{figure}
\centering
\includegraphics[angle=0,width=\columnwidth]{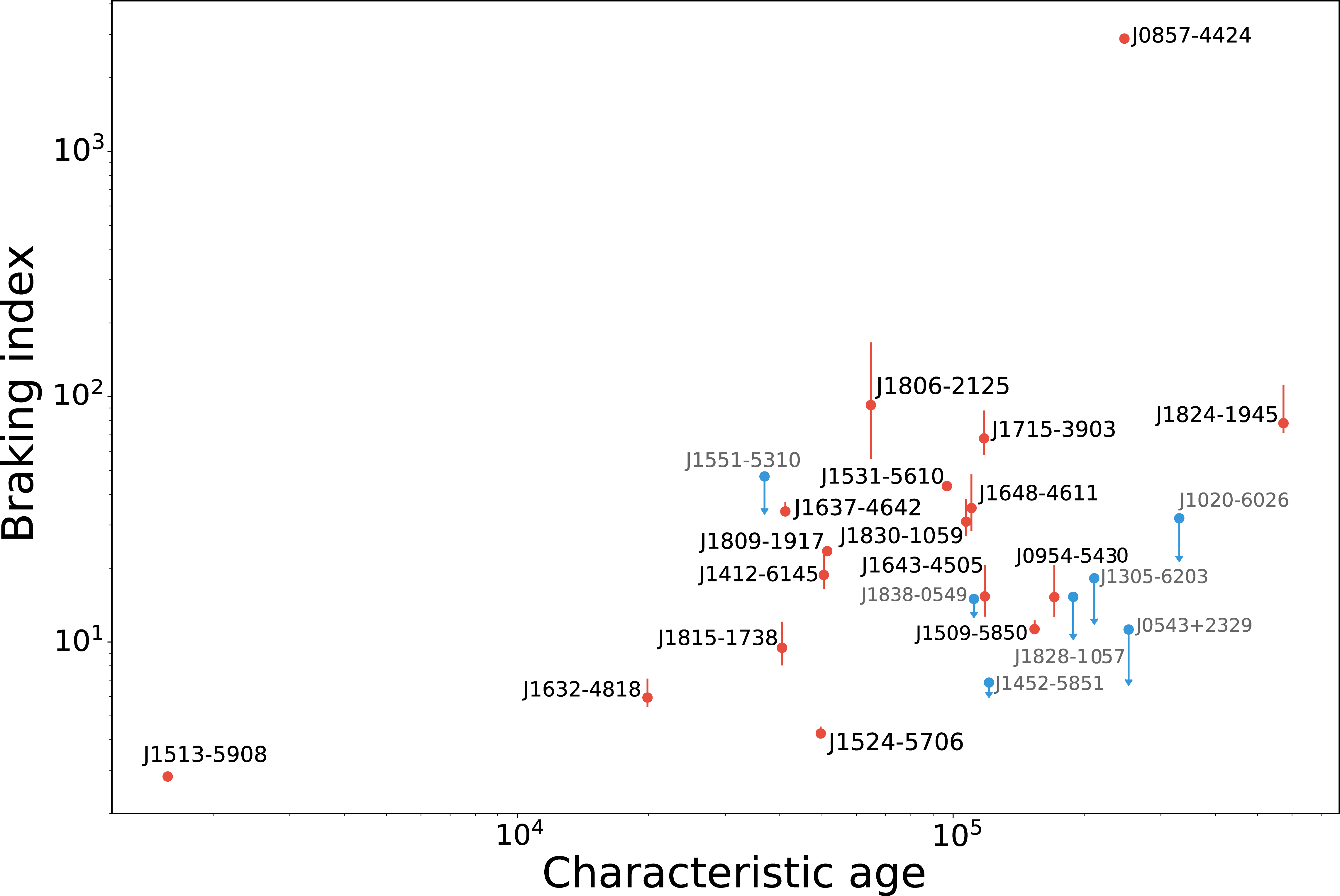}
\caption{\label{agevsbrake} Characteristic age ($\tau_{\rm c}$) versus braking index for 17 pulsars discussed in this paper (red circles) along with seven non-detections of $n$ (from PSJ19) with strongly constrained limits (blue circles). The uncertainties on the red circles are 97.5\% and 2.5\% confidence limits. The pulsars with negative values of $n$, PSRs J1738--2955 and J1833--0827 are not included in this figure. They have characteristic ages of 85~kyr and 147~kyr respectively. }
\end{figure}

We find evidence for a mild correlation (r = 0.34$\pm$0.01) between the characteristic age and $n$ as shown in Figure \ref{agevsbrake}. The correlation coefficient is computed based on the Spearmann correlation test, which is typically robust to outliers. PSR~J1513--5908 is the youngest pulsar in our sample with a characteristic age of 1.6~kyr, while PSR~J1824--1945 has a characteristic age of 570~kyr. PSR~J0857--4424 has the highest measured $n$ which is clearly an outlier for its characteristic age; this is further discussed in Section \ref{0857_wideorbit_sec}. In PSJ19, we reported limits on $n$ for pulsars for which the $\ddot{\nu}$ model was not preferred. Adding pulsars from PSJ19 with the most constrained upper limits on $n$ (represented by blue circles) does not affect this correlation. If we consider a line of constant age drawn at $\tau_{\rm c}$ $\sim$ 120 kyr in Figure \ref{ppdot_evolution}, we detect $n$ in 12 out of the 19 younger pulsars as opposed to detecting $n$ in only 7 out of 66 pulsars with $\tau_{\rm c}$ > 120 kyr. Two questions thus arise, first why do we measure significant braking indices in the younger pulsars and secondly, how can we connect the short-term braking values with the long-term evolutionary tracks in $P$-$\dot{P}$ space?

\subsection{The detectability of braking index} \label{braking_detection}
To address the first question posed above, we discuss the various factors that play a role in measuring a braking index. The detectability of $n$ for a pulsar depends on, $\nu$, $\dot{\nu}$, $\ddot{\nu}$, $T_{\rm span}$, the mean red-noise amplitude ($A_{\rm red}(\rm{TN})$) and the spectral index ($\beta(TN)$) of the timing noise.

Consider the residuals, $R(t)$, induced through a $\ddot{\nu}$:
\begin{equation}
R_{\ddot{\nu}}(t)=\frac{1}{6}\left(\frac{\ddot{\nu}}{\nu}\right) t^{3},
\end{equation}
The residuals are sampled at equal intervals over a time span $-T/2<t<T/2$. The variance of these residuals is then
\begin{equation} \label{variance_residuals}
\sigma_{\ddot{\nu}}^{2}(T) =\frac{1}{16128}\left(\frac{\ddot{\nu}}{\nu}\right)^{2} T^{6}.
\end{equation}

The variance of a power law, expressed as $P(f)=A f^{\beta}$ is 
\begin{equation} \label{variance_powerlaw}
\begin{aligned} \sigma_{\rm PL}^{2} &=\int_{1 / T}^{\infty} d f A f^{\beta} 
\\ &=\left.\frac{A}{\beta+1} f^{\beta+1}\right|_{1 / T} ^{\infty} 
\\ &=-\frac{A}{\beta+1} T^{-\beta-1} \end{aligned},
\end{equation}

where $A$ is the amplitude and $\beta$ is the spectral index. 
Relating equations \ref{variance_residuals} and \ref{variance_powerlaw},  and expressing $\ddot{\nu}$ in terms of $n$ (from equation \ref{braking_index}) we derive an expression for the power spectral density associated with $n$
\begin{equation} \label{braking_psd}
P_{n}(f)=\frac{n^{2}}{2688}\left(\frac{\dot{\nu}}{\nu}\right)^{4} f^{-7}.
\end{equation}
This can be used to assess the detectability of a cubic in the presence of other noise sources in the timing residuals. White noise, arising largely due to measurement errors, has a flat spectrum and if the amplitude of the cubic is higher than the white noise level, it will be detectable. Timing noise is a red noise process and is modelled as a power law process with a spectral index $\beta$ (see equation 4). Considering equation \ref{braking_psd} we can see that:
\begin{itemize}
    \item the power spectral density of the cubic is larger in pulsars with smaller characteristic ages (\citealt{Espinoza_Vela}) ,
    \item in pulsars where the spectral index of the timing noise is $\beta > -7$, the cubic spectral power will dominate over the timing noise for large time intervals,
    \item in pulsars where $\beta < -7$ (steep timing noise) the timing noise will dominate the cubic even with long observing time spans.
\end{itemize}

The S/N ($S$) of a deterministic signal ($r$) characterized by a noise covariance matrix ($C$) is expressed as,
\begin{equation}
S =\sqrt{\boldsymbol{r}^{\dagger} \mathbf{C}^{-1} \boldsymbol{r}}.
\end{equation}
The covariance matrix ($C$) describes the contributions from stochastic red noise and white noise components. A cubic signal is likely to be detected when $S \gg 1$. As the parameters describing the red noise process are measured directly from the data, there is an inherent degeneracy between the noise covariance matrix and the signal, which is not taken into account in this methodology. In the frequency domain, the noise covariance matrix is diagonal ($C_{ij}=0$ for $i\neq j$) and is expressed as,
\begin{equation}
C_{i i}=\left(A f_{i}^{\beta}+\frac{2 T \sigma_{WN}^{2}}{N}\right),
\end{equation}
where $\sigma_{WN}$ is the variance of the white noise components and is assumed to be equal in all observations. 

Using the noise covariance matrix and the signal, a detection statistic can be calculated to determine the S/N of a cubic in the presence of white and red noise components,

\begin{equation} \label{notional_cubic_snr}
S^{2} =\sum_{i} r_{i}^{\dagger} r_{i} C_{i i}^{-1}  =\left(\frac{n^{2}}{2688}\right)\left(\frac{\dot{\nu}}{\nu}\right)^{4} \sum_{i} \frac{f_{i}^{-7}}{A f_{i}^{\beta}+(2 T \sigma_{WN}^{2} / N)}.
\end{equation}

Using the red noise models, the time $T$ and the $\sigma_{\rm{WN}}$ for each of the pulsars in PSJ19, we can compute a $S_3$ based on equation \ref{notional_cubic_snr} for $n=3$. For 15 of the 19 pulsars with measured braking indices, we find that $S_3\gg 1$, giving a good indication of the usefulness of the metric. The remaining four pulsars have large values of $n$ which when scaled accordingly will boost the value of $S_3$.
Pulsars for which we do not detect a significant value of $n$ but have $S_3>1$ seem to favour a model with low-frequency components. The youngest pulsar in our sample without a measured $n$ is PSR~J1551--5310. It has $S_3=0.9$ due to a combination of dominant timing noise ($\beta = -7.3$) and a relatively short timing baseline of $\sim$ 7 years. The value of $S_3$ increases with time in those pulsars with shallow timing noise, hence we expect to detect more braking indices as our time span increases.

\subsection{Pulsar evolution in the $P$-$\dot{P}$ diagram.}
For PSR~J1513--5908 we stated that since our measurement of the second braking index ($m$) is consistent with the expected value from the standard spin-down model, it is reasonable to assume that $n$ has not yet evolved with time. It is thus interesting to speculate on the future evolution of PSR J1513$-$5908. If its braking index remains constant, the evolutionary track will take it above the bulk of the population in the $P$-$\dot{P}$ plane. It will cross the radio death-line with a period of $\sim$ 7 s, and perhaps then occupy the part of the $P$-$\dot{P}$ diagram containing the X-ray dim isolated neutron stars (\citealt{xins_1,xins_2}). On the other hand its braking index could increase with time, and bring it close to PSR J1632--4818 in $\sim$ 34 kyr, eventually following evolutionary tracks characterised by larger values of $n$ to become a $\sim$1~s pulsar with ever decreasing $\dot{P}$.

PSRs~J1738--2955 and J1833--0827 have a negative braking index implying that they evolve upwards and to the right in $P$-$\dot{P}$ space. A similar upward evolution is seen for PSR J1734--3333, which has a measured $n$ of 0.9$\pm$0.2 (\citealt{espinoza_magnetar_braking}) and is speculated to attain the rotational properties of a magnetar in $\sim$ 30 kyr as a consequence of surface magnetic field growth. 
PSR~J1833--0827 is at a distance of $\sim$4.3~kpc, estimated both from DM (\citealt{ymw}) and HI absorption measurements (\citealt{weisberg_hi_distance}), and has been plausibly associated with the SNR W41 (\citealt{gaensler_johnston_1995}). Our transverse velocity measurement of 800$\pm$200 km s$^{-1}$, suggests that the pulsar is indeed moving away from the SNR. We speculate that the magnetic field of the pulsar could still be increasing, thus causing the negative braking index. Another possibility is that the geometry of this pulsar is evolving towards orthogonality as is the case for the Crab pulsar (\cite{lyne_2013}). The dense plasma from the pulsar wind nebula surrounding this pulsar (\citealt{esposito_xray_pwn}) could exert a retardation torque on the neutron star causing a change in $\alpha$ (\citealt{beskin_torque_crab}). Constraints on the viewing geometry and its time dependence could prove fruitful.

What then is the link between the high values of $n$ measured here over $\sim$ 15 year time-scales with the evolution over timescales of Myr? We know of pulsars with similar characteristics to the ones discussed here, but with higher glitch rates. The spin-down of these pulsars is dominated by large braking indices ($n > 50$) between the glitches followed by a ``reset" of the $\dot{\nu}$ at the glitch itself (\citealt{yu_glitches}, \citealt{Espinoza_Vela}). \cite{Espinoza_Vela} therefore argue that long-term value of $n$ is indeed close to the canonical value of 3. Could it be then, that in our sample of pulsars, we are measuring $n$ in between glitches separated by $>20$~yr and that a future glitch will reset the $\dot{\nu}$ back to its expected value from $n=3$? Based on the theoretical work of \cite{Alpar_1993} and \cite{alpar_baykal} who postulated a weak coupling between the crust of the star and the superfluid interior, \cite{yu_glitches} demonstrated an observational relationship describing the inter-glitch value of ${\ddot{\nu}_g}$:
\begin{equation}
    \ddot{\nu}_g = 10^{-2.8\pm1.4} \,\, \dot{\nu} / T_g
\end{equation}
where $T_g$ is the mean time between glitches. In turn, \cite{fuentes_glitch} showed that $T_g \propto \dot{\nu}^{-1}$. Combining this with equation \ref{braking_index}, yields the remarkable relationship for the braking index between glitches, $n_g$, which then depends only on $\nu$,
\begin{equation}
    n_g = 10^{-0.2\pm1.4} \,\, \nu.
\end{equation}
We do not find a good correlation between $n$ and $\nu$ in our data, a formal fit yields $n_g = 10^{+0.3} \,\, \nu$ but with a large scatter. We therefore surmise that our values of $n$ contain an ``intrinsic" value higher than 3 coupled with an inter-glitch value. We also note that if the glitch resets the evolutionary track back to a slope described by  $n=3$ for long glitch intervals, then this implies that the change in $\Delta\dot{\nu} / \dot{\nu}$ must be larger than $10^{-2}$ in this set of pulsars. Therefore, there should be a correlation between $\Delta\dot{\nu} / \dot{\nu}$ and the waiting time between glitches. However, current data is too sparse to determine this. In any case, the evidence remains strong for $n$ to increase in the long-term in order to populate the $P-\dot{P}$ plane with the known pulsar population, either due to the alignment of the magnetic and rotational axes of the pulsar (\citealt{JK17}) or magnetic field decay (\citealt{vigano_bfield}).

If the braking index is indeed increasing with age, then it could imply that older pulsars have smaller inclination angles, as proposed by \cite{Tauris_Manchester_1998}, \cite{weltevrede_johnston_2008}, \cite{JK_19_pulsewidth}. Measuring the pulsar's geometry would then help to confirm this idea. For the 19 pulsars in this sample, only three pulsars have been subject to geometrical analysis (PSRs J1513--5908, J1531--5610 and J1648--4611), and even these are poorly constrained (\citealt{rookyard}). Comparisons between $n$ and $\alpha$ for a large sample of young pulsars will be an important test to further constrain theoretical predictions.

\subsection{A wide-orbit companion for PSR~J0857--4424?} \label{0857_wideorbit_sec}
A gravitational interaction between a pulsar and a companion in a wide orbit can manifest itself as the second derivative of the pulsar spin period (\citealt{matthews_wideorbit,wideorbit1,wideorbit2}). This can be expressed as (from \citealt{wideorbit1}),
\begin{equation}
\ddot{P} = \frac{G^{1 / 3} M_{\mathrm{c}} P \sin i}{\left(M_{\mathrm{c}}+M_{\mathrm{psr}}\right)^{2 / 3} c}\left(\frac{2 \pi}{P_{\mathrm{b}}}\right)^{7 / 3} \cos \phi,
\end{equation}
where $G$ is the gravitational constant, $c$ is the speed of light, $M_{\mathrm{psr}}$ is the mass of the pulsar, $M_{\mathrm{c}}$ is the mass of the orbiting companion, $P$ is the pulsar's spin period, $i$ is the orbital inclination angle, $a$ is the orbital semi-major axis and $\phi$ is the orbital phase. If we attribute the entirety of the measured $\ddot{P}$ to this effect then the orbital period ($P_{\mathrm{b}}$) is given by
\begin{equation} \label{wideorbit_eq}
P_{\mathrm{b}} = \left( \frac{M_{\rm c} \cos \phi \sin i G^{1/3} P 2\pi^{7/3}}{(M_{\rm c}+M_{\rm psr})^{2/3} \ddot{P}} \right) ^{3/7},
\end{equation}
For simplicity, if we assume a circular orbit, with $M_{\mathrm{psr}}$ of 1.44$M_{\odot}$ and $\sin i$ = 0.5, and additionally set $\cos\phi=1$, we can derive an empirical relation between $M_{\mathrm{c}}$ and the maximum $P_{\mathrm{b}}$ for a range of companion masses. 

We consider the case for PSR~J0857--4424, the pulsar with the highest value of $n$ in our sample. We take companion masses of  0.1, 1 and 10$M_{\odot}$ and compute $P_{\mathrm{b}}$ using equation \ref{wideorbit_eq}. Table \ref{wideorbit_table} reports the computed orbital periods, and the corresponding orbital separation for various companion masses. The estimated distance to PSR~J0857--4424 from the DM measurement is $\sim$ 2.8~kpc. At this distance, a 0.1$M_{\odot}$ main-sequence star will have an apparent magnitude of $\sim$ 29, while 1$M_{\odot}$ and 10$M_{\odot}$ main-sequence stars will have apparent magnitudes of $\sim$ 17 and $\sim$ 11 respectively. We examined the \textit{Gaia} database for stars within a radius of 10 arcsec from the pulsar position and found no objects (\citealt{gaia_dr2}). Since the limiting magnitude for \textit{Gaia} is 20, we can place an upper limit of $\sim$ 0.8$M_{\odot}$ on a main sequence companion. However, if the companion is a 1$M_{\odot}$ white dwarf or a neutron star, it would not be detected by \textit{Gaia}. 

It is hard to determine the likelihood of such a wide binary system existing. While we know at least one system with an orbital period in excess of 50~yr (\citealt{lyne_wideorbit}, \citealt{wideorbit_CNg}), the companion is a 15~$M_{\odot}$ main sequence star easily seen in the optical. At least one millisecond pulsar (\citealt{1620_wideorbit}) appears to have a low-mass companion in a very long orbit but this system is in a globular cluster which permits such exotic formation scenarios. Therefore we entertain a binary orbit as a potential explanation of the high braking index of PSR~J0857--4424 albeit an unlikely one.

\begin{table}
\caption{\label{wideorbit_table}
Estimated upper limits the orbital periods and orbital separations for three different companion masses for PSR J0857--4424 computed using equation \ref{wideorbit_eq}.} 
\resizebox{\columnwidth}{!}{
\begin{tabular}{lrrrrrrrrrrr}
\hline
\hline
Companion mass & Orbital period & Orbital separation  \\
($M_{\odot}$) & (yr) & (AU) \\
\hline
0.1 & 60 & 20   \\
1 & 130 & 35 \\
10 & 230 & 85  \\
\hline
\end{tabular}}
\end{table}

\section{Conclusions} \label{conclusion}
There are several strands to the work presented here. First, we restate the important point that implementing a full timing model, which includes timing noise parameterization into a Bayesian framework like {\sc temponest}, means that models can be evaluated and compared based on their Bayesian evidence. This is not possible in the generalized least-squares method implemented in {\sc tempo2}. We have shown that the values of $n$ derived in PSJ19 are robust to the addition of historical data, which emphasizes that they are unlikely to be due to unseen exponential glitches in the past. Even in the presence of glitches, successful detection and modelling of the glitch parameters maintains the consistency of $n$ across long timing baselines. Furthermore, we show that we can distinguish between models with and without an exponential glitch recovery signal and that any unseen glitch prior to our data commencing, must have little or no influence on the results presented here either because (a) the glitch size is small and/or (b) that the glitch decay time is less than 1000 days. This is supported by glitch statistics which inform that it is highly unlikely for all the pulsars presented here to have suffered large enough glitches in the immediate past to cause the presence of an exponential signal in the data.

We are therefore left with the conclusion that the large values of $n$ measured here are indeed indicative of the way a pulsar spins down, at least over a timescale of a decade or more. We consider it likely that $n$ gets reset at the time of a glitch and that the inter-glitch value of $n$ can be large. We predict a correlation between the wait time between glitches and the value of the step in $\dot{\nu}$.


Based on these results, we discuss the implications of the evolution of pulsars in the $P$-$\dot{P}$ diagram and find a moderate correlation of braking index with characteristic age, which can be attributed to either magnetic field decay and/or changes in the pulsar's inclination angle. We discuss an evolutionary path for PSR~J1513--5908 depending on the future behaviour of its braking torque and speculate on the effect of a dense PWN nebula surrounding PSR~J1833--0827 in imparting it with a negative braking index. We outline a metric to compute the detectability of $n$ in the presence of timing noise. Finally we explore the possibility of a wide-orbit companion inducing a very high braking index in PSR~J0857--4424. 

\section{Acknowledgements}
The Parkes radio telescope is part of the Australia Telescope, which is funded by the Commonwealth Government for operation as a National Facility managed by CSIRO. This work made use of the gSTAR and OzSTAR national HPC facilities. gSTAR is funded by Swinburne and the Australian Government Education Investment Fund. OzSTAR is funded by Swinburne and the National Collaborative Research Infrastructure Strategy (NCRIS). This work is supported through Australian Research Council (ARC) Centre of Excellence CE170100004. A.P thanks Chris Flynn for help with the \textit{Gaia} analysis and George Hobbs for stimulating discussions on glitches and timing noise. A.P. acknowledges support from CSIRO Astronomy and Space Science. R.M.S. acknowledges support through ARC grant CE170100004. M.B, S.O, and R.M.S. acknowledge support through ARC grant FL150100148. Work at NRL is supported by NASA. This work also made use of standard Python packages (\citealt{numpy}, \citealt{scipy}, \citealt{pandas}, \citealt{matplotlib}), Chainconsumer (\citealt{chainconsumer}) and Bokeh (\citealt{bokeh}). 




\bibliographystyle{mnras}
\bibliography{p574}




\bsp	
\label{lastpage}
\end{document}